\date{}
\documentclass[10pt,aps,pra,showpacs]{revtex4-1}
\usepackage{graphicx,psfrag,amsmath,amssymb,amsfonts,latexsym,color,dcolumn,bm}

\newcommand{\tr}{\text{tr}}
\newcommand{\xp}{x_{\parallel}}
\newcommand{\yp}{y_{\parallel}}
\newcommand{\kp}{k_{\parallel}}

\newcommand{\bkp}{\textbf{k}_\parallel}

\begin{document}
\title{Quantum friction between graphene sheets }

\author{M. Bel\' en Farias $^1$ \footnote{mbelfarias@df.uba.ar}}
\author{C\'esar D. Fosco $^2$ \footnote{fosco@cab.cnea.gov.ar}}
\author{Fernando C. Lombardo$^1$ \footnote{lombardo@df.uba.ar}}
\author{Francisco D. Mazzitelli$^{1,2}$ \footnote{fdmazzi@cab.cnea.gov.ar}}

\affiliation{$^1$ Departamento de F\'isica, FCEyN, UBA and IFIBA, CONICET,
 Pabell\'on 1, Ciudad Universitaria, 1428 Buenos Aires, Argentina}
\affiliation{$^2$ Centro At\'omico Bariloche and Instituto Balseiro,
Comisi\'on Nacional de Energ\'\i a At\'omica, 
R8402AGP Bariloche, Argentina}

\date{today}

\begin{abstract}  
We study the Casimir friction phenomenon in a system consisting of two
flat, infinite, and  parallel graphene sheets, which are coupled to the
vacuum electromagnetic (EM) field. Those couplings are implemented, in the
description we use, by means of specific terms in the effective action for
the EM field. They incorporate the distinctive properties of graphene, as
well as the relative sliding motion of the sheets.  Based on this
description, we evaluate two observables due to the same physical effect:
the probability of vacuum decay and the frictional force.  The system
exhibits a threshold for frictional effects, namely, they only exist if
the speed of the sliding motion is larger than the Fermi velocity of the
charge carriers in graphene.  
\end{abstract}
\maketitle
\section{Introduction}\label{sec:intro} 
Under propitious circumstances, quantum vacuum fluctuations
produce macroscopically observable consequences. Such is the case when a
quantum field, and hence its fluctuations, satisfy non-trivial boundary
conditions.
One of the most celebrated physical realizations of this is the Casimir
force  between two neutral bodies having non-trivial EM response
functions (which, in some cases, behave as approximate realizations of
idealized boundary conditions).
This effect has been predicted and  experimentally measured for several
different geometries~\cite{book_bordag,book_milonni,book_milton,lamoreaux2004casimir,milton2004casimir,reynaud2001quantum}.

Qualitatively different effects, also due to the vacuum fluctuations, may
arise when the bodies are set into motion or, more generally, when some
external agent renders the boundary condition(s) time-dependent. The
resulting effect may involve dissipation and, when the boundary conditions
experiment non-vanishing accelerations, \textit{real} photons can be
excited out of the quantum vacuum.  This embodies the most frequently
considered version of the so called dynamical Casimir effect (DCE)
\cite{review_dyncas}, also known as `motion induced radiation'.  

A more startling situation appears when a purely quantum, dissipative,
frictional force arises between bodies moving with {\em constant\/}
relative speed.  Here, the effect is due to the quantum degrees of freedom,
living on the moving media, which are excited out of the vacuum, while the EM
field is nevertheless required as a mediator for those fluctuations.  The
resulting effect, termed {\em Casimir friction} has been extensively 
studied and some of the issues involved in its calculation have spurred
some debate~\cite{review_friction,volokitin_persson,pendry97,pendry_debate}.

 We recall that Casimir friction predictions have been obtained mostly 
for dielectric materials. In this paper, we study the same effect, but for two graphene sheets.  We
argue that graphene has unusual properties which render its theoretical
study more interesting. Indeed,  because of graphene's low dimensionality
and particular crystalline structure, its low-energy excitations behave as
massless Dirac fermions (with the Fermi velocity $v_F$ playing the role of
light's speed). This yields an unusual semi-metallic
behavior~\cite{rev_graphene}, as well as peculiar transport and optical
properties~\cite{opacgraphene,opacexpgraphene, farias2013}.  

In natural units (which we adopt here) the mass dimensions of the response
function of graphene in momentum space can only be given by 
the momentum itself. Indeed, the only other ingredients: $v_F$ and the
effective electric charge of the fermions, are dimensionless.  And, when a sheet is
moving at a constant speed $v$, another dimensionless object, $v$ itself,
enters into the game (see below).  Thus the non-trivial dependence of the
macroscopic, Casimir friction observables,  will exhibit the remarkable
property of being a function of $v$ and $v_F$, the overall (trivial)
dimensions of the respective magnitude being determined purely by geometry:
size and distance between
sheets, like in the static Casimir effect between perfect mirrors.  

A somewhat related but different effect, also termed  `quantum friction', has 
been studied for graphene in Ref. \cite{volokitin2011quantum}. Note, however,
that in that work the system consists of a single static graphene sheet 
over an SiO 2 substrate. The frictional force acts, in this case, on graphene’s
charge carriers, which are assumed to have a constant drift velocity $v$
with respect to the substrate.

In our study below, we start from a consideration of the microscopic model
for two graphene sheets coupled to the EM field. Those microscopic degrees
of freedom correspond to Dirac fields in $2+1$ dimensions which, in a
functional integral formulation, are integrated out. That integration, plus
the free gauge field action, produces an in-out effective action for the
latter. Integrating the gauge field, we finally get an effective action for
the full system, the imaginary part of which accounts for the dissipative
effects in the system,  a procedure we have followed in our previous
works~\cite{farias_friction,farias2016}. 

We perform our calculations within a functional integral
formalism~\cite{fosco2007_moving_mirrors,fosco2011_sidewise}, and after
evaluating the probability of vacuum decay, we relate  the imaginary part
of the in-out effective action to the frictional force on the
plates, and plot the latter as a function of the velocity $v$.  

The structure of this paper is as follows: in Section~\ref{sec:themodel}, we
introduce the microscopic model considered in this article. Then we derive
an `effective action' for the EM field, namely, an Euclidean action which,
in our description, is a functional of $A_\mu$, the gauge field
corresponding to the vacuum EM field. In order to achieve that, we need to
find the form of the vacuum polarization tensor for moving graphene (as seen 
from rest) assuming relativistic effects can be
neglected. 

In Section~\ref{sec:effective}, we calculate the full effective action resulting
from the integration of the EM field. That effective action, when rotated
to Minkowski space, is applied to the calculation of the probability of
vacuum decay, as a function of the velocity of the sliding graphene sheet. 
In Section~\ref{sec:potencia}, we relate the imaginary part of the in-out
effective action to the dissipated power, and thereby to the frictional
force on the moving plate.
Section~\ref{sec:conclu} contains our conclusions.
\section{The model}\label{sec:themodel}
We first introduce the Euclidean action ${\mathcal S}$, for the EM field
plus the two graphene sheets, one of them static, the other moving at a
constant velocity (which is assumed to be parallel to the sheets).  The
action depends on the gauge field and on the Dirac fields, the latter confined to
the mirrors. ${\mathcal S}$ naturally decomposes into three terms: 
\begin{equation}\label{eq:defs}
{\mathcal S}[A; \bar\psi, \psi] \;=\; {\mathcal S}^{(0)}_{\rm g}[A] \,+\,
{\mathcal S}^{(0)}_{\rm d}[\bar\psi, \psi] \,+\, {\mathcal S}^{({\rm
int})}_{\rm dg}[\bar\psi, \psi,A] \;, 
\end{equation}
where ${\mathcal S}^{(0)}_{\rm g}$ is the free (i.e., empty-space) action
for the EM field:
\begin{equation}\label{eq:defsg0}
	{\mathcal S}^{(0)}_{\rm g}[A]\;=\; \frac{1}{4} \int d^4x \, F_{\mu\nu} F_{\mu\nu} \;,
\end{equation}
with $F_{\mu\nu} = \partial_\mu A_\nu - \partial_\nu A_\mu$, 
whilst ${\mathcal S}^{(0)}_{\rm d}$ and ${\mathcal S}_{\rm d g}^{({\rm
int})}$ are the actions for the free Dirac matter fields and for their
interactions with the gauge field, respectively. Indices from the middle of
the Greek alphabet ($\mu, \nu, \ldots$) run from $0$ to $3$, with $x_0
\equiv c t$.

Both ${\mathcal S}^{(0)}_{\rm d}$ and ${\mathcal S}_{\rm d g}^{({\rm
int})}$ are localized on the regions occupied by the two sheets, which we
denote by $L$ and $R$ (each letter will be used to denote both a
mirror and the spatial region it occupies). Our choice of Cartesian
coordinates is such that $L$ is defined by $x_3=0$ and $R$ by $x_3=a$. 
We adopt conventions such that $\hbar=c=1$.

We introduce $\Gamma$, the effective action for the full system defined
in (\ref{eq:defs}) by ${\mathcal S}$. It can be written in
terms of ${\mathcal Z}$, the zero-temperature partition function, which may
be represented as a functional integral: 
\begin{equation}\label{eq:defgamma}
e^{-\Gamma} \;\equiv\; \mathcal{Z} \;\equiv\;\int [{\mathcal D} A]
{\mathcal D}\bar\psi {\mathcal D}\psi e^{-{\mathcal S}[A; \bar\psi,\psi]}
\;, 
\end{equation}
where $[{\mathcal D} A]$ is the gauge field functional integration measure
including gauge fixing.

The effect of the Dirac fields on the gauge field is taken into
account by integrating out the former, we introduce ${\mathcal
S}_{\text{eff}}$, as follows:
\begin{equation}\label{eq:defseffa}
e^{- {\mathcal S}_{\text{eff}}[A]} \;\equiv\; \int {\mathcal D}\bar\psi
{\mathcal D}\psi \, e^{-{\mathcal S}[A; \bar\psi,\psi]} \;, 
\end{equation}
so that:
\begin{equation}\label{eq:defgamma1}
e^{-\Gamma} \;\equiv\;\int [{\mathcal D} A] \, e^{-{\mathcal
S}_{\text{eff}}[A]}\;. 
\end{equation}

Recalling our previous discussion and conventions, we write the effective action as
\begin{equation}
	{\mathcal S}_{\text{eff}}[A] \;=\;  {\mathcal S}^{(0)}_{\rm g}[A] \,+\, 
	{\mathcal S}^{(\text{int})}_{\rm g}[A] \;.
\end{equation}
In the next two subsections, we deal with the determination of ${\mathcal
S}^{(\text{int})}_{\rm g}$, which is the result of the integration of the
fermionic degrees of freedom.

\subsection{Effective action contribution due to the static sheet}
\label{sec:statmir}
As in~\cite{dynamical_matter}, the effective interaction term for the 
gauge field in the presence of graphene sheets stems from two essentially
$2+1$ dimensional theories, coupled to the $3+1$ dimensional gauge field. 
Therefore, ${\mathcal S}^{(\text{int})}_{\rm g} = {\mathcal
S}^{(\text{L})}_{\rm g} + {\mathcal S}^{(\text{R})}_{\rm g}$, where each
term is due to the respective plate. The fact that one of the
sheets is moving is irrelevant to the dimensionality of those theories,
since the surface it occupies is invariant under the sliding motion.

Let us first consider ${\mathcal S}^{(\text{L})}_{\rm g}[A]$, due to the
static sheet at $x_3 = 0$. Up to the quadratic order in the gauge field,
following~\cite{dynamical_matter}, we write such contribution as follows:
\begin{equation}\label{eq:defpil}
{\mathcal S}^{(\text{L})}_{\rm g}[A] \,=\,
\frac{1}{2} \,\int d^3x_\parallel \int d^3y_\parallel \,
A_\alpha(x_\parallel,0) \, \Pi_{\alpha\beta}(x_\parallel,y_\parallel) \,
A_\beta(y_\parallel,0) \;,
\end{equation}
where indices from the beginning of the Greek alphabet ($\alpha, \beta,
\ldots$) are assumed to take the values $0, 1, 2$ and are used here to
label spacetime coordinates on the $2+1$ dimensional world-volume of each
sheet. Those coordinates have been denoted collectively by $x_\parallel$.
Regarding the corresponding $2+1$ dimensional Fourier momentum, we use $\kp
\equiv (k_0,k_1, k_2)$, and ${\mathbf k_\shortparallel} \equiv
(k_1,k_2)$ for its spatial part.

The tensor kernel $\Pi_{\alpha\beta}$ is the vacuum polarization tensor
(VPT) for the matter field on the $L$ plane.  Under the assumptions of
time-independence, as well as invariance under spatial rotations and
translations, this  tensor can be conveniently decomposed in Fourier space
into orthogonal projectors.  Indeed, since it has to verify the
Ward identity:
\begin{equation}\label{eq:trans}
	k_\alpha \widetilde{\Pi}_{\alpha \beta}(k)\;=\; 0 \;,
\end{equation}
(the tilde is used to denote Fourier transformation) the irreducible
tensors (projectors) along which
$\widetilde{\Pi}_{\alpha\beta}$ may be decomposed must satisfy the
condition above and may be constructed using as building blocks the
objects: $\delta_{\alpha\beta}$, $k_\alpha$, and \mbox{$n_\alpha =
(1,0,0)$}.  By performing simple combinations among them, we also
introduce: \mbox{$\breve{k}_\alpha \equiv k_\alpha - k_0 n_\alpha$}, and
\mbox{$\breve{\delta}_{\alpha\beta} \equiv \delta_{\alpha\beta} - n_\alpha
n_\beta$}. 

Since we cannot guarantee that the VPT will be proportional to 
${\mathcal P}^\perp_{\alpha\beta} \,\equiv\, \delta_{\alpha\beta}- \frac{k_\alpha
k_\beta}{k^2}$, we construct two independent tensors satisfying the transversality
condition (\ref{eq:trans}), ${\mathcal P}^t$ and ${\mathcal P}^l$, defined
as follows:
\begin{equation}
{\mathcal P}^t_{\alpha\beta} \,\equiv\, \breve{\delta}_{\alpha\beta} -
\frac{\breve{k}_\alpha\breve{k}_\beta}{\breve{k}^2}
\end{equation}
and 
\begin{equation}
{\mathcal P}^l_{\alpha\beta} \,\equiv\, {\mathcal
P}^\perp_{\alpha\beta}\,-\, {\mathcal P}^t_{\alpha\beta} \;.
\end{equation}
Defining also:
\begin{equation}
{\mathcal P}^\shortparallel_{\alpha\beta} \,\equiv\, \frac{k_\alpha
k_\beta}{k^2} \;,
\end{equation}
we verify the algebraic properties:
	$$
	{\mathcal P}^\perp + {\mathcal P}^\shortparallel = I \;,\;\;
	{\mathcal P}^t + {\mathcal P}^l = {\mathcal P}^\perp 
	$$
	$$
	{\mathcal P}^t {\mathcal P}^l = {\mathcal P}^l {\mathcal P}^t = 0
	\;,\;\; {\mathcal P}^\shortparallel {\mathcal P}^t = {\mathcal P}^t {\mathcal P}^\shortparallel = 0 
	\;,$$
	$$\;\; {\mathcal P}^\shortparallel {\mathcal P}^l = {\mathcal P}^l {\mathcal P}^\shortparallel = 0 \;,
	$$
\begin{equation}
\big({\mathcal P}^\perp\big)^2 = {\mathcal P}^\perp \;,
\big({\mathcal P}^\shortparallel\big)^2 = {\mathcal P}^\shortparallel \;,
\big({\mathcal P}^t\big)^2 = {\mathcal P}^t \;,
\big({\mathcal P}^l\big)^2 = {\mathcal P}^l \;.
\end{equation}

Note that $\delta_{\alpha\beta}$, ${\mathcal P}^\perp_{\alpha\beta}$,
and ${\mathcal P}^\shortparallel_{\alpha\beta}$ are second order Lorentz
tensors. 
The other projectors,  ${\mathcal P}^t$ and ${\mathcal P}^l$ are not: they
explicitly single out the time-like coordinate in their definition. On the
other hand, Lorentz tensors will tend to Galilean ones in the low speed
limit.

For a  general medium, one has:
\begin{equation}
	\widetilde{\Pi}_{\alpha\beta}(\kp) \,=\,  
	g_t\big(k_0, {\mathbf k}_\parallel\big) \, {\mathcal P}^t_{\alpha\beta} \,+\, 
	g_l\big(k_0, {\mathbf k}_\parallel\big) \, {\mathcal P}^l_{\alpha\beta}
	\;,
	\label{piscalar}
\end{equation}
where $g_t$ and $g_l$ are model-dependent scalar functions.

If the matter-field action were relativistic, we would have $g_t=g_l\equiv
g$,  a scalar function of $\kp$, and the VPT would be proportional to a single projector:
\begin{equation}
{\widetilde \Pi}_{\alpha \beta}(k_\parallel) \;=\; g(k_\parallel)
\, {\mathcal P}^\perp_{\alpha\beta} \; .
\end{equation}

On the other hand, for the case of graphene,  we may present the well-known
results for its VPT \cite{rev_graphene}, as follows:
\begin{equation}
	\widetilde{\Pi}_{\alpha\beta}(\kp) =\frac{e^2 N |m|}{4\pi}  
	F\big(\frac{k_0^2 + v_F^2 {\bkp}^2}{4 m^2}\big) 
	\Big[ {\mathcal P}^t_{\alpha\beta} + 
	\frac{k_0^2+{\bkp}^2}{k_0^2+ v_F^2 {\bkp}^2} {\mathcal P}^l_{\alpha\beta} 
	\Big]
\end{equation}
where:
\begin{equation}
	F(x) \;=\;   1 - \frac{1 - x}{\sqrt{x}} \, \arcsin[ ( 1 +
	x^{-1})^{-\frac{1}{2}}]\;,  
\end{equation}
$m$ is the mass (gap), $N$ the number of $2$-component Dirac fermion
fields, and $v_F$ the Fermi velocity (in units where $c=1$).

Here we will consider gapless graphene ($m=0$) and define
$\alpha_N \equiv \frac{e^2 N}{16}$, so that
\begin{align}
	\widetilde{\Pi}_{\alpha\beta} &= \alpha_N  \, 
	\sqrt{k_0^2 + v_F^2 {\bkp}^2}\; 
	\Big[ {\mathcal P}^t_{\alpha\beta} \,+\, 
	\frac{k_0^2+{\bkp}^2}{k_0^2+ v_F^2 {\bkp}^2} {\mathcal P}^l_{\alpha\beta} 
	\Big] \nonumber\\
                         &= \alpha_N \, \sqrt{k_0^2+ {\bkp}^2} \,  
	\Big[ \, \sqrt{\frac{k_0^2 + v_F^2 {\bkp}^2}{k_0^2 +
	{\bkp}^2}} \;
	{\mathcal P}^t_{\alpha\beta}  + 
	\sqrt{\frac{k_0^2 + {\bkp}^2}{k_0^2 + v_F^2
	{\bkp}^2}} \; {\mathcal P}^l_{\alpha\beta} 
	\Big] \;.
\end{align}
 We see explicitly that the mass
dimension of the VPT is given by the momentum, as mentioned in the Introduction. 

We conclude the discussion on ${\mathcal S}_{\text{g}}^{(\text{L})}$ (we
work at the second order in the coupling constant) by writing it in a $3+1$
dimensional looking form,
\begin{equation}
\label{eq:seffl}
{\mathcal S}_{\text{g}}^{(\text{L})}[A] \;=\;  \frac{1}{2}\,  \int d^4x
\int d^4y \;  A_\alpha(x) V^{({\text L})}_{\alpha\beta}(x,y) A_\beta(y) \, ,
\end{equation}
where $V^{(\text{L})}_{\alpha\beta}(x,y)$ is given by
\begin{equation}
\label{eq:vl}
V^{(\text{L})}_{\alpha\beta}(x,y) \;=\; \delta(x_3) \,
\Pi_{\alpha\beta}(x_\parallel,y_\parallel) \, \delta(y_3) \;.
\end{equation}

\subsection{Effective action due to the moving graphene sheet}
We already know the expression for the effective action due to the static
mirror at $x_3=0$; let us see now how to derive from it the corresponding object for
the moving sheet at $x_3=a$. We assume that its constant velocity is much smaller than
$c$, so that the form it adopts in two different inertial systems may be
derived using Galilean transformations. Besides, the material media
descriptions are usually restricted to the same regime, namely, small
speeds with respect to the laboratory system (the response functions are
usually defined in a comoving system). 

Since we need to write the effective action in one and the same system, we
need to write the gauge field appearing in ${\mathcal S}_{\text{g}}^{(\text{R})}$
in the laboratory system, the one used in the previous subsection. We also need to
refer them to the same choice of coordinates. Thus,
\begin{equation}
\label{eq:seffr}
{\mathcal S}_{\text{g}}^{(\text{R})}[A] \;=\;  \frac{1}{2}\,  \int d^4x
\int d^4y \; A_\alpha(x) V^{(\text{R})}_{\alpha\beta}(x,y) A_\beta(y) \, ,
\end{equation}
where 
\begin{equation}
\label{eq:rpot}
V^{(\text{R})}_{\alpha\beta}(x,y) \;=\; \delta(x_3 - a)
\Pi'_{\alpha\beta}(x_\parallel,y_\parallel) \delta(y_3 - a) \;,
\end{equation}
where the prime in an object denotes its form in the comoving system. To
write the expression above more explicitly, we need to introduce the
transformations $x_\parallel' = \Lambda(v) x_\parallel$, where
$x_\parallel$ is the column vector $x_\parallel = \left(\begin{array}{c}
x_0\\ x_1\\ x_2 \end{array}\right)$. 

Those transformations can be obtained by
keeping the first non-trivial term in an expansion in powers of $v$. Since
we have adopted conventions such that $c=1$, and our metric is Euclidean,
we see that:
\begin{equation}
\Lambda(v)=\left(\begin{array}{c c c c }
1 & v & 0 \\
-v & 1 & 0 \\
0 & 0& 1
\end{array}
\right) \, 
\end{equation}
(i.e., they are rotation matrices expanded for small angles). We have only
kept the three spacetime coordinates corresponding to the sheets, since the
role of the $x_3$ coordinate is irrelevant here.
Note that the matrix includes the transformation of the time coordinate, while
Galilean transformations do not include that transformation and are given by:
\begin{equation}
\Lambda_G(v)=\left(\begin{array}{c c c c }
1 & 0 & 0 \\
-v & 1 & 0 \\
0 & 0& 1
\end{array}
\right) \,. 
\end{equation}

The EM field, on the other hand, transforms as
$A'_\alpha(x')=\Lambda_{\alpha\beta} A_\beta(x)$.
Regarding the VPT, we have
\begin{equation}
\Pi'_{\alpha\beta}(\xp ', \yp ') \;=\;\Lambda_{\alpha\gamma}
\Lambda_{\beta\delta} \, \Pi_{\gamma\delta}(\xp, \yp) \,. 
\end{equation}
Thus,
\begin{equation}
\Pi'_{\alpha\beta}(\xp, \yp) \;=\;\Lambda_{\alpha\gamma}
\Lambda_{\beta\delta} \, \Pi_{\gamma\delta}(\Lambda^{-1}\xp,\Lambda^{-1}\yp) \,. 
\end{equation}
Then we see that:
\begin{equation}
\label{eq:pot}
V^{(\text{R})}_{\alpha\beta}(x,y) \;=\; \delta(x_3 - a)
\,\Lambda_{\alpha\gamma}
\Lambda_{\beta\delta} \, \Pi_{\gamma\delta}(\Lambda^{-1}\xp,\Lambda^{-1}\yp)
\, \delta(y_3 - a) \;,
\end{equation}

In momentum space, we can write 
\begin{eqnarray}\label{eq:aux1}
	\widetilde{\Pi}'_{\alpha\beta}(k_\parallel) &=&
	\Lambda_{\alpha\gamma} \Lambda_{\beta\delta} \,
	\widetilde{\Pi}_{\gamma\delta}(\Lambda^{-1} \kp) \nonumber\\
	&=&
	\Lambda_{\alpha\gamma} \Lambda_{\beta\delta} \,
	\widetilde{\Pi}_{\gamma\delta}(k_0 - v k_1, k_1 + v k_0, k_2) \;.
\end{eqnarray}

\subsection{The full effective action ${\mathcal S}_{\text
g}^{(\text{int})}$ for graphene}
Putting together the previous results, we have that:
\begin{equation}
\label{eq:seffint}
{\mathcal S}_{\text g}^{(\text{int})}\;=\;
\frac{1}{2}\,  \int d^4x
\int d^4y \; A_\alpha(x) \big[ V^{(\text{L})}_{\alpha\beta}(x,y) +
V^{(\text{R})}_{\alpha\beta}(x,y) \big] A_\beta(y) \;,
\end{equation}
or
\begin{equation}
{\mathcal S}_{\text g}^{(\text{int})}[A] \;=\; \frac{1}{2} \, \int
dx_3 \int dy_3 \int \frac{d^3\kp}{(2\pi)^3} \,
\tilde{A}^*_\alpha(\kp,x_3) \left[
	\delta(x_3) \tilde{\Pi}_{\alpha\beta}(\kp) \delta(y_3) + \delta(x_3 - a)
\tilde{\Pi}'_{\alpha\beta}(\kp) \delta(y_3-a) \right]
\tilde{A}_\beta(\kp,y_3) \,,
\end{equation}
with $\tilde{\Pi}'_{\alpha\beta}(\kp)$ as defined in (\ref{eq:aux1}).

Let us now study in more detail the form of
$\widetilde{\Pi}'_{\alpha\beta}(k_\parallel)$. 
We have:
\begin{equation}
\widetilde{\Pi}'_{\alpha\beta}(\kp) \,=\,   g_t\big(\Lambda^{-1}
	k_\parallel\big) \,{\mathcal P}'^t_{\alpha\beta}
\,+\,  g_l\big(\Lambda^{-1}
	k_\parallel\big) \,{\mathcal P}'^l_{\alpha\beta} \;.
\end{equation}	
Now, we will see that the two projectors remain invariant under Galilean
transformations.
Indeed, we first note that the Lorentz projectors, which enter into the
definition of the Galilean ones, are indeed invariant (we
use the approximate Lorentz form for the transformation matrix): 
\begin{align}
	{\mathcal P}'^\perp_{\alpha\beta}(k_\parallel)
	&=\,\Lambda_{\alpha\gamma} \,
\Lambda_{\beta\delta} {\mathcal P}^\perp_{\gamma\delta}(\Lambda^{-1}
k_\parallel) = {\mathcal P}^\perp_{\alpha\beta}(k_\parallel) \\
{\mathcal P}'^\shortparallel_{\alpha\beta}(k_\parallel) &=\, 
\Lambda_{\alpha\gamma} \, \Lambda_{\beta\delta} \,
{\mathcal P}^\shortparallel_{\gamma\delta}(\Lambda^{-1} k_\parallel)
={\mathcal P}^\shortparallel_{\alpha\beta} (k_\parallel)\;,
\end{align}
while for the Galilean tensor $P^t$ we verify explicitly that:
\begin{equation}
	{\mathcal P}'^t_{\alpha\beta}(k_\parallel)
	\,=\,(\Lambda_G)_{\alpha\gamma} \,
	(\Lambda_G)_{\beta\delta} {\mathcal P}^t_{\gamma\delta}((\Lambda_G)^{-1}
k_\parallel) = {\mathcal P}^t_{\alpha\beta}(k_\parallel) \;.
\end{equation}

Since ${\mathcal P}^l$ is defined in terms of the previously considered
three projectors, we see that: 
\begin{equation}
	{\mathcal P}'^l_{\alpha\beta} = {\mathcal P}^l_{\alpha\beta} \;.
\end{equation}
Thus, we conclude that:
\begin{equation}
\widetilde{\Pi}'_{\alpha\beta}(k_\parallel) \,=\,   g_t\big(\Lambda^{-1}
	k_\parallel\big) \,{\mathcal P}^t_{\alpha\beta}
\,+\,  g_l\big(\Lambda^{-1}
	k_\parallel\big) \,{\mathcal P}^l_{\alpha\beta} \;.
\end{equation}

We are interested in small relative velocities between the plates, so we are able to use the simpler expression
\begin{equation}
\widetilde{\Pi}'_{\alpha\beta}(\kp) \,=\,   g_t\big(k_0-v k_1, k_1 + v k_0, k_2\big) \,{\mathcal P}^t_{\alpha\beta}
\,+\,  g_l\big(k_0-v k_1, k_1 + v k_0, k_2 \big) \,{\mathcal P}^l_{\alpha\beta} \;.
\end{equation}
where
\begin{align}
\label{eq:gs}
g_t(k_\parallel)&=\alpha_N \sqrt{k_0^2 + v_F^2 \textbf{k}_\parallel ^2} \\
g_l(k_\parallel)&=\alpha_N \sqrt{k_0^2 + v_F^2 \textbf{k}_\parallel ^2} \, \frac{k_0^2 + \bkp ^2}{k_0^2+ v_F^2\bkp ^2}\nonumber
\end{align}
\section{Effective action}\label{sec:effective}
With all the previous considerations, we are now in a position to write the
total action for the gauge field, containing the effective influence of the
graphene plates. In Fourier space:
\begin{equation}
S_{\text{g}}[A] \,=\, \frac{1}{2} \int dx_3 \int dy_3 \int \frac{d^3\kp}{(2\pi)^3} \tilde{A}^*_\alpha (\kp, x_3) M_{\alpha\beta} (\kp,x_3,y_3) \tilde{A}_\beta(\kp,y_3)
\end{equation}
where the kernel $M_{\alpha\beta} (\kp,x_3,y_3)$ can be written as
\begin{equation}
M_{\alpha\beta} (\kp,x_3,y_3)=M^0_{\alpha \beta}(\kp,x_3,y_3)+M^{\text{int}}_{\alpha\beta} (\kp,x_3,y_3)
\end{equation}
where $M^0$ is the free kernel for the vacuum EM field
\begin{equation}
M^0_{\alpha \beta}(\kp,x_3,y_3)=-\partial_3^2 \delta(x_3-y_3) \mathcal{P}^\parallel_{\alpha \beta} + (-\partial_3^2+\kp^2) \delta(x_3-y_3) \left[ \mathcal{P}^l_{\alpha\beta}+\mathcal{P}^t_{\alpha\beta} \right] \, ,
\end{equation}
and $M^{\text{int}}$ contains the effective interaction with the plates' internal degrees of freedom
\begin{equation}
M^{\text{int}}_{\alpha\beta} (\kp,x_3,y_3)= \tilde{V}^{(\text{L})}_{\alpha\beta}(\kp,x_3,y_3)+\tilde{V}^{(\text{R})}_{\alpha\beta}(\kp,x_3,y_3) \, .
\end{equation}

The generating functional for the system is defined by
\begin{equation}
\mathcal{Z}=\int \left[ \mathcal{D}A\right] e^{-S_{\text g}[A]}
\end{equation}
where $\left[ \mathcal{D}A\right]$ is gauge-fixed. Formally, it is equivalent to writting
\begin{equation}
\mathcal{Z}=\left[ \det \left( M_{\alpha \beta}(\kp,x_3,y_3)\right)\right]^{-\frac{1}{2}}
\end{equation}

Now, since we have chosen a complete set of projectors $\left\lbrace \mathcal{P}^\parallel, \mathcal{P}^t, \mathcal{P}^l \right\rbrace$, we can uniquely decompose the gauge field in their directions $\tilde{A}_\alpha \equiv \tilde{A}^\parallel_\alpha + \tilde{A}^t_\alpha + \tilde{A}^l_\alpha$, thus writing the functional integral over $A$ as three independent functional integrals
\begin{equation}
[ \mathcal{D}\tilde{A} ] = \mathcal{D }\tilde{A}^\parallel \,  \mathcal{D} \tilde{A}^t \, \mathcal{D} \tilde{A}^l \, .
\end{equation}
This means that the integrating functional for the system can be written as the direct product of three independent integrating functionals:
\begin{equation}
\mathcal{Z}=\big[ \det \big( M^\parallel (\kp,x_3,y_3) \big)\big]^{-\frac{1}{2}} \left[ \det \left( M^t (\kp,x_3,y_3)\right)\right]^{-\frac{1}{2}} \left[ \det \left( M^l(\kp,x_3,y_3)\right)\right]^{-\frac{1}{2}} \equiv \mathcal{Z}^\parallel \mathcal{Z}^t \mathcal{Z}^l
\end{equation}
where we have defined the kernels:
\begin{equation}
\label{eq:mpar}
M^\parallel (\kp, x_3, y_3) = - \partial_3^2 \delta(x_3-y_3) \mathcal{P}^\parallel \, ,
\end{equation}
\begin{align}
\label{eq:ml}
M^l(\kp,x_3,y_3)=&\big\lbrace (-\partial_3^2 + \kp^2) \delta(x_3-y_3) + g_l (k_0,k_1,k_2) \delta(x_3) \delta(y_3)   \nonumber \\
&+    g_l (k_0-v k_1,k_1+v k_0,k_2) \delta(x_3-a) \delta(y_3-a ) \big\rbrace \mathcal{P}^l \, ,
\end{align}
and
\begin{align}
\label{eq:mt}
M^t(\kp,x_3,y_3)=&\big\lbrace (-\partial_3^2 + \kp^2) \delta(x_3-y_3) + g_t (k_0,k_1,k_2) \delta(x_3) \delta(y_3)   \nonumber \\
&+    g_t (k_0-v k_1,k_1+v k_0,k_2) \delta(x_3-a) \delta(y_3-a ) \big\rbrace \mathcal{P}^t \, .
\end{align}

Given Eq. \eqref{eq:mpar}, it is easy to see that $\mathcal{Z}^\parallel$   is a free contribution that does not account for the presence of the plates. It is thus simply a normalization factor, and we shall not take it into account in the following. The remaining factors $\mathcal{Z}^t$ and $\mathcal{Z}^l$ are formally equivalent but different, except for relativistic materials.

Regarding the effective action, it is easy to see that it shall have two independent contributions
\begin{equation}
\Gamma \equiv \Gamma^t + \Gamma^l = \frac{1}{2} \tr \log M^t + \frac{1}{2} \tr \log M^l \, .
\end{equation}

We shall now work out the formal expression for $\Gamma^t$; the
corresponding expression for $\Gamma^l$ is obtained by the substitutions
$g_t \to g_l$, $\mathcal{P}^t \to \mathcal{P}^l$. As in previous works
\citep{farias_friction,fosco2011_sidewise}, we will perform a  perturbative
expansion in
the coupling constant, $e \ll 1$,  and keep only the lowest-order
non-trivial term. 

Explicitly taking
the trace over all discrete and continuous indices in this term we get a $T
\Sigma$ global factor, $T$ denoting the elapsed time and $\Sigma$ the sheets' area (this
is a reflection of the time and (parallel) space translation invariances of
the system).  Since $\Gamma^t$ is extensive in those magnitudes, we
work instead with  $\gamma^t \equiv \frac{\Gamma}{T \Sigma}$, which is
given by 
\begin{equation}
	\gamma^t = -\frac{  1}{4} \int \frac{d^3\kp}{(2\pi)^3} \int dx_3
	\int dy_3 \int du_3 \int dv_3 G_{\alpha \gamma}(\kp,x_3,y_3)
	V^t_{\gamma \delta}(\kp,y_3,u_3) G_{\delta \beta}(\kp,u_3,v_3)
	V^t_{\beta \alpha} (\kp, v_3, x_3) \;.
\end{equation}
Here, $G_{\alpha \gamma}(\kp,x_3,y_3)$ denotes the respective components
of the free Euclidean propagator for the gauge field, and we have
introduced
\begin{equation}
V^t \equiv  \left[ g_t (k_0,k_1,k_2) \delta(x_3) \delta(y_3)  +   g_t
(k_0-v k_1,k_1+v k_0,k_2) \delta(x_3-a) \delta(y_3-a ) \right]
\mathcal{P}^t  \;.
\end{equation}
We only consider in what follows the `crossed' terms, namely, those
involving both $g_t (k_0,k_1,k_2)$ and $g_t (k_0-v k_1,k_1+v k_0,k_2)$,
since they are the only ones that lead to friction (the others can be shown
to be $v$-independent).

Taking into account that, in the Feynman gauge: 
\begin{equation}
G_{\alpha \beta}(\kp,x_3,y_3) \equiv \delta_{\alpha\beta} G(\kp,x_3,y_3) =
\delta_{\alpha\beta} \int \frac{dk_3}{2\pi} \frac{e^{i k_3 (x_3 -
y_3)}}{\kp^2 + k_3^2} \,,
\end{equation}
and the properties of the projectors, we see that
\begin{equation}
	\gamma^t \,=\, -\frac{1}{2} \int \frac{d^3\kp}{(2\pi)^3} G(\kp,a,0) G(\kp,0,a)
g_t(k_0,k_1,k_2) g_t(k_0-v k_1, k_1+v k_0,k_2) \,.
\end{equation}
The procedure and outcome for the $\Gamma^l$ contribution are entirely
analogous, thus we may write ($s=t,l$):
\begin{equation}
\label{eq:effacts}
\gamma^s \;=\; -\frac{1}{2} \int \frac{d^3\kp}{(2\pi)^3} \; G(\kp,a,0) G(\kp,0,a)
g_s(k_0,k_1,k_2) g_s(k_0-v k_1, k_1+v k_0,k_2) \, .
\end{equation}
Thus,
\begin{equation}
\label{eq:gammas}
	\gamma^s \;=\; -\frac{1}{8\,  a^3} \int \frac{d^3\kp}{(2\pi)^3} \;
	\frac{e^{-2\sqrt{k_0^2 + k_1^2 + k_2^2}}}{k_0^2 + k_1^2 + k_2^2}
	g_s(k_0,k_1,k_2) g_s(k_0-v k_1, k_1+v k_0,k_2) \, ,
\end{equation}
where we have rescaled the momenta $ak_\alpha\to k_\alpha$ in order to factorize the dependence of the effective action
with the distance between sheets. Note that $\gamma_s$ is the effective action per unit time and area, and therefore has units
of $(length)^{-3}$.

Before evaluating the imaginary part of the real time (in-out) effective action, we would like to stress that the Euclidean  effective action
$\gamma$, when evaluated at $v=0$,  gives the usual Casimir interaction energy per unit area $E_C$ between the graphene sheets. 
As described in the Appendix A, the result is 
  \begin{equation}
  \label{eq:casfor}
E_\text{C} \approx - \frac{\alpha_N^2}{128 \pi}  \frac{1}{a^3} \frac{1}{v_F} \, .
\end{equation}
 As expected,  due to the absence of dimensionful constants in the microscopic description of graphene,   the Casimir energy has the 
usual $1/a^3$ dependence of the static vacuum interaction energy for perfect conductors. {Eq. \eqref{eq:casfor} is quadratic in the coupling constant $\alpha_N$, while the Casimir force found in \cite{bordag_casimir_graphene} is linear. The reason is that
we calculate the force between two graphene plates,
while in \cite{bordag_casimir_graphene} the interaction 
between a perfect conductor and a graphene sheet is considered.} 
 
\subsection{Imaginary part of the effective action}
In order to compute the imaginary part of the in-out effective action, we have to rotate the Euclidean result to real time.
To that end,  we will rewrite each contribution in a way that simplifies the forthcoming 
discussion. Note that we can write the two functions $g_t$ and $g_l$ as follows:
\begin{align}
g_t(k_\parallel)&= \alpha_N \int_{-\infty}^{+\infty}\frac{dk_3}{\pi} 
\frac{k_0^2 + v_F^2 \textbf{k}_\parallel^2}{k_0^2 + v_F^2 \textbf{k}_\parallel ^2 + k_3^2} \\
g_l(k_\parallel)&=\alpha_N \int_{-\infty}^{+\infty} \frac{dk_3}{\pi} 
\frac{k_0^2 + \textbf{k}_\parallel ^2}{k_0^2 + v_F^2 \textbf{k}_\parallel ^2 + k_3^2} \;.
\end{align}
Then we see that,
\begin{equation}
\gamma^t = -\frac{ \alpha_N^2 }{8\, a^3}
\int \frac{dk_3}{\pi}\int\frac{dp_3}{\pi} \int
\frac{d^3k_\parallel}{(2\pi)^3} \,\frac{e^{-2\sqrt{k_0^2 + k_1^2 +
k_2^2}}}{k_0^2 + k_1^2 + k_2^2} 
\frac{k_0^2 + v_F^2 \textbf{k}_\parallel ^2}{k_0^2 + v_F^2 \textbf{k}_\parallel ^2 + k_3^2}
\frac{(k_0-k_1 v)^2 + v_F^2 [(k_1+k_0 v)^2 + k_2^2]}{(k_0-k_1 v)^2 + v_F^2 [(k_1+k_0 v)^2 + k_2^2] + p_3^2}
 \,,
\end{equation}
and
\begin{equation}
\gamma^l = -\frac{\alpha_N^2}{8\, a^3}  
\int \frac{dk_3}{\pi}\int\frac{dp_3}{\pi} \int
\frac{d^3k_\parallel}{(2\pi)^3} \, \frac{e^{-2\sqrt{k_0^2 + k_1^2 +
k_2^2}}}{k_0^2 + k_1^2 + k_2^2} 
\frac{k_0^2 +\textbf{k}_\parallel^2}{k_0^2 + v_F^2 \textbf{k}_\parallel ^2
+ k_3^2} \frac{k_0^2 +\textbf{k}_\parallel^2}{(k_0-k_1 v)^2 + v_F^2
[(k_1+k_0 v)^2 + k_2^2] + p_3^2} \,.
\end{equation}

In real-time, the longitudinal contribution to the effective action is
\begin{equation}
\label{eq:inoutl}
\gamma^l = \frac{i \alpha_N^2}{8\, a^3} \,
\int \frac{dk_3}{\pi} \int \frac{dp_3}{\pi} \int\frac{d^3\kp}{(2\pi)^3}
\frac{e^{2 i \sqrt{k_0^2 - \bkp^2 + i \epsilon}}}{k_0^2 - \bkp^2 + i
\epsilon} \times
\frac{k_0^2 - \textbf{k}_\parallel^2}{k_0^2 - v_F^2 \textbf{k}_\parallel ^2 - k_3^2 + i \epsilon} \times
\frac{k_0^2 - \textbf{k}_\parallel^2}{(k_0-k_1 v)^2 - v_F^2 [(k_1-k_0 v)^2 + k_2^2] - p_3^2 + i \epsilon}
 \,.
\end{equation}

We shall be concerned first with the integral along $k_0$, which may be
first conveniently written as follows:
\begin{equation}
\int_0^\infty dk_0 \frac{e^{2i \sqrt{k_0^2-\bkp^2+i\epsilon}}}{k_0^2-\bkp^2+i\epsilon} \left[ f_1(k_0) f_2(k_0)+f_1(-k_0) f_2(-k_0)\right]
\end{equation}
where
\begin{align}
f_1(k_0)&\equiv \frac{k_0^2 - \textbf{k}_\parallel^2}{k_0^2 - v_F^2 \textbf{k}_\parallel ^2 - k_3^2+ i \epsilon}\\
f_2(k_0)&\equiv \frac{k_0^2 -\textbf{k}_\parallel^2}{(k_0-k_1 v)^2 - v_F^2 [(k_1-k_0 v)^2 + k_2^2] - p_3^2 + i \epsilon} \nonumber \; .
\end{align}

In order to perform this integral, we proceed along a similar line to the
one followed in \cite{farias_friction}, namely, to study the analytical structure
of the functions $f_1$ and $f_2$ in order to perform a Wick-rotation by
means of a Cauchy integration on the quarter of a circle located in the first
quadrant. Note that the rest of the integrand is the same as the
one dealt with in~\cite{farias_friction}: it presents two branch-cuts and
two poles, none of them in the first quadrant, hence they do not contribute to
the Cauchy integral. 
Let us then consider the poles of $f_1(k_0)=f_1(-k_0)$; they are located at:
\begin{equation}
k_0=\pm \sqrt{v_F^2 \bkp^2+k_3^2-i \epsilon}\approx \pm \sqrt{v_F^2 \bkp^2+k_3^2} \mp \frac{i \epsilon}{2 \sqrt{v_F^2 \bkp^2+k_3^2}} \, .
\end{equation}
Since none of them is located in the first quadrant, they will not
contribute to the Cauchy integral either. For the $f_2(k_0)$ function,
they are located at:
\begin{equation}
k_0^{(\pm)} = \frac{1}{(1-v_F^2 v^2)} \left\lbrace v k_1 (1 - v_F^2) \pm \sqrt{v^2 k_1^2(1-v_F^2)^2+(1-v_F^2v^2)\left[(v_F^2 - v^2) k_1^2 + v_F^2 k_2^2 + p_3^2 - i \epsilon \right]}\right\rbrace
\end{equation}

It can be seen that only $k_0^{(-)}$ may have a positive imaginary
part (and thus be located in the first quadrant).  We shall denote its
position by $\Lambda_A\equiv k_0^{(-)}$.
The condition for it to belong to the first quadrant is
$\text{Re}\Lambda_A > 0$. We first note that,  if $k_1 < 0$, then
$\text{Re}\Lambda_A < 0$ and there is no pole located on the first
quadrant. On the other hand, for positive values of $k_1$, one can show
that:
\begin{align*}
\text{Re}\Lambda_A  >0  \Leftrightarrow -(v_F^2 - v^2) k_1^2 - (v_F^2 k_2^2 + p_3^2) >0 \, .
\end{align*}
Clearly, when $v<v_F$, the LHS of the last equation is negative-definite, and
the inequality can never be fulfilled. Hence, for velocities smaller than
the Fermi velocity of the material, this pole can never be located in the
first quadrant. Finally, when $v>v_F$, we will have a pole in the first
quadrant when
\begin{equation}
k_1 > \sqrt{\frac{v_F^2 k_2^2 + p_3^2}{v^2-v_F^2}} \, .
\end{equation}

Proceeding in a completely analogous way for the $f_2(-k_0)$ term, one can also check
that just one pole may belong to the
first quadrant when $v>v_F$. The position of that pole is given by:
\begin{equation}
\Lambda_B=\Lambda_A- 2 v k_1\frac{1-v_F^2}{1-v_F^2v^2}\, .
\end{equation}
The pole is located on the first quadrant for momenta such that 
\begin{equation}
k_1< - \sqrt{\frac{v_F^2 k_2^2 + p_3^2}{v^2-v_F^2}}\, .
\end{equation}

Based on the previous analysis, we are now ready to perform the
Cauchy-integral along the quarter of a circle, in a rather similar fashion
as we did in~\cite{farias_friction}. The result is:
\begin{align}
\label{eq:antesdejuntarAB}
\gamma^l =& \frac{i \alpha_N^2}{8\, a^3 (2 \pi)^3} \int \frac{dk_3}{\pi}\int  \frac{dp_3}{\pi} \int dk_2 \int dk_1 \left\lbrace -i\int_0^\infty dp_0 \frac{e^{-2\kp}}{\kp^2} f_1(i p_0) \left[  f_2(i p_0) + f_2(-ip_0) \right] \right. \nonumber\\ 
&+ 2 \left. \pi i \theta(v-v_F) \left[ \theta\left(k_1- \sqrt{\frac{v_F^2 k_2^2 + p_3^2}{v^2-v_F^2}}\right)\text{Res}(F_A(k_0),\Lambda_A) + \theta\left(-k_1- \sqrt{\frac{v_F^2 k_2^2 + p_3^2}{v^2-v_F^2}}\right)\text{Res}(F_B(k_0),\Lambda_B)  \right] \right\rbrace \, ,
\end{align}
where
\begin{equation}
F_A(k_0)=F_B(-k_0)= \frac{e^{2i \sqrt{k_0^2-\bkp^2+i \epsilon}}}{k_0^2-\bkp^2+i \epsilon} f_1(k_0) f_2(k_0) \, .
\end{equation}

Since we are interested in computing the dissipative effects on the system, we shall take the imaginary part of the effective action. It is easy to see that $f_1(p_0) \in \mathbb{R}$ and that $f_2(i p_0) + f_2(-i p_0) \in \mathbb{R}$ also. Hence, the imaginary part of the longitudinal contribution to the effective action will be given by
\begin{align}
	\text{Im} \gamma^l = -\frac{\alpha_N^2}{16 \pi^2\, a^3} \,\theta(v-v_F)
	 \int
\frac{dk_3}{\pi} \int \frac{dp_3}{\pi} \int dk_2
\int dk_1\text{Im} &\left\lbrace  \, \text{Res}(F_A(k_0),\Lambda_A)  \,
\theta\left(k_1- \sqrt{\frac{v_F^2 k_2^2 + p_3^2}{v^2-v_F^2}}\right)
\right\rbrace  \, .
\end{align}
From this equation, we see that there is no longitudinal
contribution to the quantum friction for plates moving with a relative
velocity smaller than the Fermi velocity of the material.\\

Regarding the transversal contribution to the effective action, let us
first rotate it back to real time:
\begin{equation}
\label{eq:inoutt}
\gamma^t = \frac{i \alpha_N^2}{8\, a^3} \,
\int \frac{dk_3}{\pi}\int \frac{dp_3}{\pi}\int \frac{ d^3\kp}{(2\pi)^3}\frac{e^{2 i \sqrt{k_0^2 - \bkp^2 + i \epsilon}}}{k_0^2 - \bkp^2 + i \epsilon}  \times
\frac{k_0^2 - v_F^2 \textbf{k}_\parallel^2}{k_0^2 - v_F^2 \textbf{k}_\parallel ^2 - k_3^2 + i \epsilon} \times
\frac{(k_0-k_1 v)^2 - v_F^2 (k_1-k_0 v)^2 - v_F^2 k_2^2}{(k_0-k_1 v)^2 - v_F^2 [(k_1-k_0 v)^2 + k_2^2] - p_3^2 + i \epsilon}
 \,.
\end{equation}
%
The calculation is entirely similar to the previous case. 
The imaginary part of the transversal
contribution to the in-out effective action reads
\begin{align}\label{eq:resultt}
\text{Im} \gamma^t = -\frac{\alpha_N^2}{16 \pi^2\, a^3} \,\theta(v-v_F) \int \frac{dk_3}{\pi} \int \frac{dp_3}{\pi} \int dk_2 dk_1 \text{Im} &\left\lbrace \text{Res}(F_C(k_0),\Lambda_A)  \, \theta\left(k_1- \sqrt{\frac{v_F^2 k_2^2 + p_3^2}{v^2-v_F^2}}\right)
\right\rbrace  \, ,
\end{align}
with
\begin{equation}
F_C(k_0)= \frac{e^{2i\sqrt{k_0^2-\bkp^2+i \epsilon}}}{k_0^2-\bkp^2+i \epsilon} f_3(k_0) f_4(k_0) \, ,
\end{equation}
and
\begin{align}
f_3(k_0)&=\frac{k_0^2  - v_F^2 \textbf{k}_\parallel^2}{k_0^2 - v_F^2 \textbf{k}_\parallel ^2 - k_3^2 + i\epsilon}\\
f_4(k_0)&=\frac{(k_0-k_1 v)^2 - v_F^2 (k_1-k_0 v)^2 - v_F^2 k_2^2}{(k_0-k_1 v)^2 - v_F^2 [(k_1-k_0 v)^2 + k_2^2] - p_3^2 + i \epsilon} \nonumber \, .
\end{align}
Hence we arrive to the important conclusion that  there will not
be quantum friction between two graphene plates unless they move  at a
relative velocity larger than the Fermi velocity of the internal
excitations in graphene. Note that a velocity threshold effect has also been shown to appear in dielectric materials~\cite{cherenkov}, as a consequence of a different, Cerenkov-like effect. \\

%
 The remaining integrals and the limit $\epsilon \rightarrow 0$, needed to obtain the imaginary part of the effective action,
 can be performed with some analytical and numerical calculations that are detailed in the Appendix \ref{ap1}.
The results are shown in Fig.~\ref{fig:plots}, where it may be seen that the
transverse contribution is much smaller than the longitudinal one. Then, the first plot on Fig. \ref{fig:plots}
shows the behavior of the leading contribution to the  imaginary part of the effective action as a
function of the relative velocity $v$, for a Fermi velocity of $v_F=0.003$.

\begin{figure}[h]
\includegraphics[scale=2]{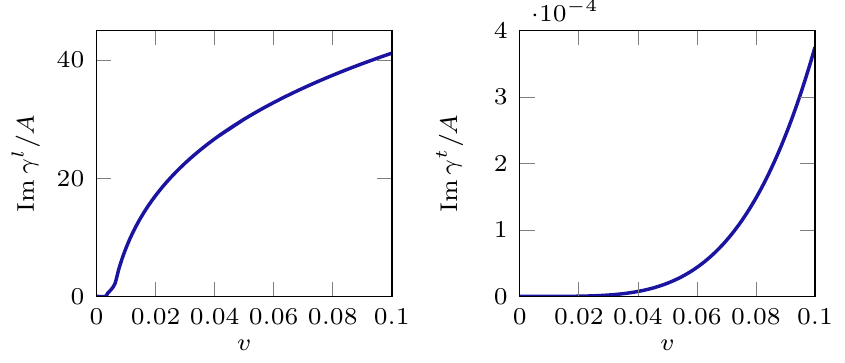} 
\caption{\label{fig:plots} Imaginary part of the effective action per unit of time and area, as a function of the relative velocity of the plate, for a typical graphene Fermi velocity $v_F=0.003$, in units of $A= \frac{\alpha_N^2}{32 \pi^2} \frac{1}{a^3}$.}
\end{figure}
%

\section{Frictional force}\label{sec:potencia}

In order to quantify the dissipation, a rather convenient observable is the
dissipated power (and its related dissipative force). Let us see how
that power is related to the imaginary part of the in-out effective action.

Dissipation arises here when the Dirac vacuum becomes unstable against the
production of a 
real (i.e., on shell) fermion pair.
The probability $\mathcal P$ of
such an event, during the whole history of the plates,  is related to the
effective action by 
\begin{equation}
\label{eq:gammaP}
2 \text{Im} \Gamma =\mathcal{P} = T \int d^3 \kp p(\kp)  \, ,
\end{equation}
where $p(\kp)$ is the probability per unit time of creating a pair of fermions on the plates with total momentum $\kp$. The result is 
proportional to the whole time elapsed
$T$,  since we are in a stationary regime (we assume this time to be a very long one after the mirror was set to motion).  Note that $\kp$ is the three-momentum injected on the system by the external conditions, i.e. 
the motion of the R-mirror.
The explicit expression for $p(\kp)$ can be read from Eqs. (\ref{eq:inoutl}) and (\ref{eq:inoutt}). It can be written as
\begin{equation}
p(\kp)=\int dk_3 \int dp_3\,  \delta(k_0-\Lambda_A) h(\kp,k_3,p_3) \, ,
\end{equation}
for some function $h$. The presence of the $\delta$-function highlights the fact that the integration in the $k_0$-complex plane
captures the contribution of a single pole at $k_0=\Lambda_{A}$.

On the other hand,  the total energy $E$ accumulated in the plates due to the excitation of the internal degrees of freedom is given by
\begin{equation}
E = T \int d^3\kp \vert k_0\vert p(\kp)\, .
\end{equation}
This energy is provided  by the external source that keeps the plate moving at a constant velocity, against the frictional force (per unit area) $F_{\text{fr}}$. The  energy balance is 
\begin{equation}
\frac{E}{T\Sigma}=v F_\text{fr}\, .
\end{equation}

\begin{figure}[t]
\includegraphics[scale=2]{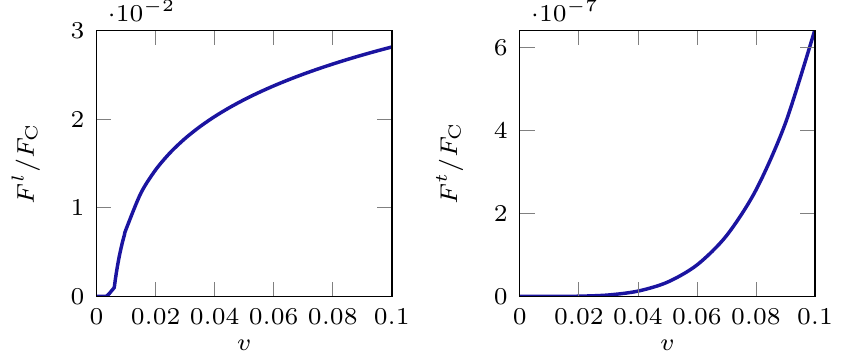} 
\caption{\label{fig:fza} Modulus of the transversal and longitudinal contributions to the force per unit of area $F_\text{fr}$ acting on the plate as a function of its relative velocity, for a typical graphene Fermi velocity $v_F=0.003$. The force is normalized by the static Casimir force between the plates.}
\end{figure}

From the reasoning above,  it is easy to see
that in order to obtain the dissipated power we can simply insert $\left| k_0 \right|$
in Eqs. \eqref{eq:inoutl} and \eqref{eq:inoutt}, repeat the procedure of 
the last section, and multiply the result by $2/v$. Note that the insertion of $\left| k_0 \right|$does not spoil the discussion
about the position of the poles, that remains unchanged. The results for the longitudinal and transverse contributions to the force are shown in Fig.
\ref{fig:fza}. We plot the frictional force normalized by the static Casimir force between the graphene sheets $F_C$, given in Eq. \eqref{eq:casfor}.

\begin{figure}[h!]
\includegraphics[scale=2]{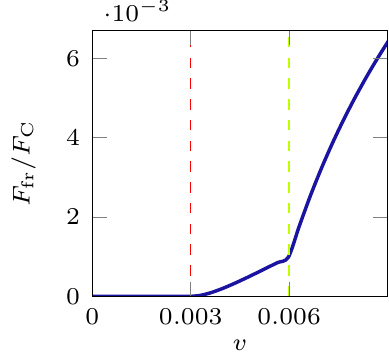} 
\caption{\label{fig:plotszoom} Modulus of the force per unit of area acting on the plate as a function of its relative velocity, for velocities close to  the Fermi velocity of graphene, $v_F=0.003$. The force is normalized by the static Casimir force between the plates.}
\end{figure}

In Fig. \ref{fig:plotszoom} we show the force for velocities close to the
Fermi velocity. There, it can be seen that the system undergoes
three different regimes regarding dissipation. For $v<v_F$, as already mentioned,  there are no
dissipative effects on the system, and the total frictional force vanishes.
For velocities $v_F<v<2 v_F$,  a frictional
force appears, but it grows comparatively slow with the velocity. For
$v>2 v_F$, however, the frictional force starts growing rapidly when the
velocity increases. 

The existence of a threshold may be justified as follows.  Let us consider  the momentum and energy balance in a small time
interval $\delta t$, assuming that both the frictional force and the
dissipated energy are driven by pair creation. 
The only relevant component of the total momentum ${\mathbf P}$ of the pair
for the (momentum) balance is the one along the direction of the velocity
$v$. Relating that component of ${\mathbf P}$ to the frictional
force, we see that: 
\begin{equation}\label{eq:bmom}
F_\text{fr} \, \delta t \; = \;  P_x \;.
\end{equation}
On the other hand, the energy balance reads
\begin{equation}\label{eq:bener}
F_\text{fr} \, v \, \delta t \; = \; {\mathcal E}
\end{equation}
where ${\mathcal E}$ is the energy of the pair. But, since the fermions are both
on-shell, we have 
\begin{equation}\label{eq:disper}
{\mathcal E} \geq v_F |P_x| 
\end{equation}
(the equal sign corresponds to a pair with momentum along the direction of
$v$).  Dividing Eq.(\ref{eq:bener}) and Eq.(\ref{eq:bmom}), and taking into account
Eq.(\ref{eq:disper}), we see that a necessary condition for friction to
happen is:
\begin{equation}
v \; \geq \; v_F \;.
\end{equation}

\section{Conclusions}\label{sec:conclu}

In this paper we computed the vacuum friction between graphene sheets subjected to a sidewise motion with constant relative velocity. The interaction between the $2+1$ Dirac fields in the graphene  sheets and the electromagnetic field has been taken into account using the known results for the comoving vacuum polarization tensor, properly transformed to the laboratory system in the case of the moving sheet. We have seen that this interaction generates an imaginary part in the effective action, that in the nonrelativistic limit can be interpreted as due to the excitation of the internal degrees of freedom produced by the relative motion between sheets.   Therefore, dissipation effect arises due to the fact the Dirac vacuum becomes unstable against the production of a real (i.e., on shell) fermion pair. We also computed the frictional force between plates using a slight modification of the calculation of the imaginary part of the effective action. 

The results for the imaginary part of the effective action and for the frictional force show an interesting phenomenon: there is a threshold for quantum friction effects, that is, there is no quantum friction when the relative velocity between sheets is smaller than the Fermi velocity. We have presented a simple argument that justifies the existence of this threshold.

The frictional force computed in this paper is much smaller than the usual Casimir force between graphene sheets, which in turn is smaller than the Casimir force between perfect conductors (at least when considering gapped graphene, see  Ref. \cite{bordag_casimir_graphene}). However, one may envisage situations in which the frictional force could be more relevant. Indeed, it has been pointed out that, at high temperatures, the Casimir force between a graphene sheet and a perfect conductor becomes comparable with that between perfect conductors \cite{fialkovsky2011finite}.  Moreover, doping can strongly enhance the Casimir force between graphene sheets \cite{bordag2016enhanced}. It would be of interest to generalize the results of the present paper to compute the frictional force in those situations, and discuss whether the enhancement of the Casimir force have a corresponding effect in the frictional force or not.

{
On the other hand, we have found that the frictional force vanishes identically for speeds smaller than $v_F$.
From the point of view of applications, graphene has been regarded as one of the most promising new materials, both for its electronic and mechanical properties. Our results imply, for example, that when graphene is used in a micro-mechanical device, Casimir friction, and its concomitant energy dissipation, will not be present below the threshold, which presumably will be the best scenario for most applications.
}

\section*{Acknowledgements}
This work was supported by ANPCyT, CONICET, UBA and UNCuyo. M.B.F would like to thank Tom\'as S. Bortol\'in for valuable insights and discussions.

\begin{appendix}

\section{Static Casimir force between two graphene sheets}\label{ap2}

In absence of dissipative effects (i.e., for $v=0$), the Euclidean vacuum persistence amplitude is
\begin{equation}
\mathcal{Z}=e^{-E_0 T}
\end{equation}
where $T$ is the elapsed time, and $E_0$ is the zero-point energy of the EM field. This means that the Casimir energy per unit of area $E_C=E_0/\Sigma$ can be obtained from the Euclidean effective action of the plates when their relative velocity vanishes, that is $E_C=\gamma_{\text{Eucl}}(v=0)$. 

Taking $v=0$ in Eq. \eqref{eq:gammas}, and recalling the definitions of $g_t$ and $g_l$ of Eq. \eqref{eq:gs}, the transversal contribution to the zero-point energy results 
\begin{align}
E_C^t=-\frac{1}{8\, a^3} \int \frac{d^3\kp}{(2 \pi)^3}    \frac{e^{-2 \sqrt{k_0^2 + \kp^2}}}{k_0^2+\kp^2} g_t^2(k_0,\kp)=-\frac{1}{48}\frac{\alpha_N^2}{(2\pi)^2} \frac{1}{a^3} (1+2 v_F^2) \,.
\end{align}
Analogously, the longitudinal contribution is given by
\begin{align}
E_C^l=-\frac{1}{8\, a^3} \int \frac{d^3\kp}{(2 \pi)^3}    \frac{e^{-2 \sqrt{k_0^2 + \kp^2}}}{k_0^2+\kp^2} g_l^2(k_0,\kp)=-\frac{1}{16}\frac{\alpha_N^2}{(2\pi)^2} \frac{1}{a^3} \frac{\arccos(v_F)}{v_F \sqrt{1-v_F^2}} \, .
\end{align}

Considering that typical Fermi velocities are much smaller than the velocity of light, the leading contribution to the static Casimir energy between two graphene sheets comes from the longitudinal effective action and reads
  \begin{equation}
E_\text{C} \approx - \frac{\alpha_N^2}{128 \pi}  \frac{1}{a^3} \frac{1}{v_F} \, .
\end{equation}%
Therefore the Casimir attractive force acting on the sheets results
\begin{equation}
\label{eq:fzacas}
F_\text{C}\approx \frac{3 \alpha_N^2}{128 \pi}  \frac{1}{a^4} \frac{1}{v_F} \, .
\end{equation}

\section{Details of the calculation of the imaginary part of the effective action}\label{ap1}

In order to obtain a final expression for the imaginary part of the
effective action, it is necessary to compute the desired residues. We will repeat the proceadure we did in \cite{farias_friction}. Let us start
with the longitudinal part:
\begin{align}
\label{eq:res}
\text{Res}(F_A(k_0),\Lambda_A)\equiv & \lim_{k_0 \rightarrow \Lambda_A} (k_0-\Lambda_A) F_A(k_0) \nonumber \\
=&\frac{e^{2i \sqrt{\Lambda_A^2-\bkp^2+i \epsilon}}}{\Lambda_A^2-\bkp^2+i \epsilon}\times \frac{\Lambda_A^2 - \textbf{k}_\parallel^2}{\Lambda_A^2 - v_F^2 \textbf{k}_\parallel ^2 - k_3^2+i \epsilon} \times \frac{\Lambda_A^2 -\textbf{k}_\parallel^2}{-2 \sqrt{v_F^2 k_1^2 (1-v^2)^2 +(1-v_F^2 v^2)(v_F^2 k_2^2 + p_3^2)}} \, ,
\end{align}
where in the last factor we have explicitly used the fact that the denominator is positive-definite and thus the limit $\epsilon \rightarrow 0$ can be taken with no further harm.

It could be shown that $\Lambda_A^2-\bkp^2$ is definite-negative in all the
integration region (that is, for $k_1>\sqrt{v_F^2 k_2^2 +p_3^2 /(v^2
-v_F^2)}$). This can be easily seen when
explicitly taking both $v$ and $v_F\ll1$, but the relation still holds for arbitrary
values of $v,v_F<1$. This means that we can set $\epsilon
= 0$ everywhere except in the second factor of \eqref{eq:res}, that can be written as:
%
%
%
\begin{equation}
\frac{1}{\Lambda_A^2 - v_F^2 \textbf{k}_\parallel ^2 - k_3^2+i \epsilon}=\frac{1}{g(k_1)+i \epsilon} \, ,
\end{equation}
with
\begin{equation}
g(k_1,k_2,k_3,p_3)=\Lambda_A^2 - v_F^2 \textbf{k}_\parallel ^2 - k_3^2 \, .
\end{equation}
Now we can explicitly take the limit $\epsilon \rightarrow 0$,
\begin{equation}
\frac{1}{g(k_1,k_2,k_3,p_3)+i \epsilon} \rightarrow \text{p.v.} \left(\frac{1}{g} \right) - i \pi \delta \left( g(k_1,k_2,k_3,p_3) \right) \, .
\end{equation}

Therefore, the longitudinal contribution to the imaginary part of the effective action reads
\begin{align}
\text{Im} \gamma^{l} = &  \, \frac{\alpha_N^2}{32 \pi\, a^3} \,\theta(v-v_F) \int \frac{dk_3}{\pi} \int \frac{dp_3}{\pi} \int dk_1\int dk_2  \delta \left( g(k_1,k_2,k_3,p_3 \right) \theta\left(k_1- \sqrt{\frac{v_F^2 k_2^2 + p_3^2}{v^2-v_F^2}}\right)  \\
 & \times e^{-2 \sqrt{\bkp^2-\Lambda_A^2}} \frac{\textbf{k}_\parallel^2-\Lambda_A^2}{\sqrt{v_F^2 k_1^2 (1-v^2)^2 +(1-v_F^2 v^2)(v_F^2 k_2^2 + p_3^2)}} \,  .
\end{align}
Note that we have is a 4-dimensional integration of a function multiplied by the Dirac-delta function composed with $g$:
\begin{equation}
\int d^4 \kappa \mathcal{F}(\bm{\kappa}) \delta(g(\bm{\kappa})) = \intop_{\mathcal{S}/g(\bm{\kappa})=0} d\sigma \frac{ \mathcal{F}(\bm{\kappa})}{|\nabla g(\bm{\kappa})|} \, ,
\end{equation}
where in our case $\bm{\kappa}=(k_1,k_2,k_3,p_3)$. The second hand is an
integration over the 3-dimensional surface defined by
$g(k_1,k_2,k_3,p_3)=0$. We can think this surface as the one defined by the
equations $k_1=x_1(k_2,k_3,p_3)$ and $k_1=x_2(k_2,k_3,p_3)$, with
\begin{align}
x_1(k_2,k_3,p_3)&= \sqrt{\frac{u(k_2,k_3,p_3)-2 \sqrt{w(k_2,k_3,p_3)}}{v^2 \left(v_F^2-1\right)^2 \left(v v_F^2+v-2 v_F\right) \left(v v_F^2+v+2 v_F\right)}} \\
x_2(k_2,k_3,p_3)&= \sqrt{\frac{u(k_2,k_3,p_3)+2 \sqrt{w(k_2,k_3,p_3)}}{v^2 \left(v_F^2-1\right)^2 \left(v v_F^2+v-2 v_F\right) \left(v v_F^2+v+2 v_F\right)}}
\end{align}
where
\begin{align}
\label{eq:ceros}
u(k_2,k_3,p_3)=&v^2 \left(1-v_F^2\right) \left\lbrace p_3^2 \left(v_F^2+1\right) + k_2^2 v_F^2 \left[v_F^2 \left(v^2 \left(v_F^2+1\right)-2\right)+2\right]+k_3^2 \left[1+ v_F^2 \left(v^2 \left(v_F^2+1\right)-3\right)\right]\right\rbrace \nonumber \\
w(k_2,k_3,p_2)=&v^2 \left(1-v_F^2\right)^2 \left\lbrace k_3^4 v_F^2 \left(v^2-1\right)^2 + k_3^2 \left[k_2^2 v^2 v_F^2 \left(2 \left(v^2-2\right) +v_F^4+1\right)+p_3^2 \left(v^2 \left(v_F^4+1\right)-2 v_F^2\right)\right] \right. \nonumber \\
& + \left. k_2^4 v^2 v_F^4 \left[1+\left(v^2-2\right) v_F^2+v_F^4\right]+k_2^2 p_3^2 v^2 v_F^2 \left(v_F^4+1\right)+p_3^4 v_F^2 \right\rbrace \, .
\end{align}

Then we have
\begin{align}
\text{Im} \gamma^{l} = &  \frac{\alpha_N^2}{32 \pi\, a^3} \,\theta(v-v_F)  \int \frac{dk_3}{\pi} \int \frac{dp_3}{\pi} \int dk_2 \int dk_1 \sum_{i=1,2} \frac{\delta \left( k_1 - x_i \right)}{\left| \nabla g(k_1,k_2,k_3,p_3)\right|_{k_1=x_i}} \,  \theta\left(k_1- \sqrt{\frac{v_F^2 k_2^2 + p_3^2}{v^2-v_F^2}}\right)  \\
 & \times e^{-2 \sqrt{\bkp^2-\Lambda_A^2}} \frac{\textbf{k}_\parallel^2-\Lambda_A^2}{\sqrt{v_F^2 k_1^2 (1-v^2)^2 +(1-v_F^2 v^2)(v_F^2 k_2^2 + p_3^2)}} \,  .
\end{align}
The result of the integration over $k_1$ can be written as a Heaviside
step-function of a rather involved expression depending on the rest of the
integration variables. This, and the remaining integrals have been
performed numerically.

The calculation of $\gamma^t$ proceeds in a similar way.

\end{appendix}



\begin{thebibliography}{25}%
\makeatletter
\providecommand \@ifxundefined [1]{%
 \@ifx{#1\undefined}
}%
\providecommand \@ifnum [1]{%
 \ifnum #1\expandafter \@firstoftwo
 \else \expandafter \@secondoftwo
 \fi
}%
\providecommand \@ifx [1]{%
 \ifx #1\expandafter \@firstoftwo
 \else \expandafter \@secondoftwo
 \fi
}%
\providecommand \natexlab [1]{#1}%
\providecommand \enquote  [1]{``#1''}%
\providecommand \bibnamefont  [1]{#1}%
\providecommand \bibfnamefont [1]{#1}%
\providecommand \citenamefont [1]{#1}%
\providecommand \href@noop [0]{\@secondoftwo}%
\providecommand \href [0]{\begingroup \@sanitize@url \@href}%
\providecommand \@href[1]{\@@startlink{#1}\@@href}%
\providecommand \@@href[1]{\endgroup#1\@@endlink}%
\providecommand \@sanitize@url [0]{\catcode `\\12\catcode `\$12\catcode
  `\&12\catcode `\#12\catcode `\^12\catcode `\_12\catcode `\%12\relax}%
\providecommand \@@startlink[1]{}%
\providecommand \@@endlink[0]{}%
\providecommand \url  [0]{\begingroup\@sanitize@url \@url }%
\providecommand \@url [1]{\endgroup\@href {#1}{\urlprefix }}%
\providecommand \urlprefix  [0]{URL }%
\providecommand \Eprint [0]{\href }%
\providecommand \doibase [0]{http://dx.doi.org/}%
\providecommand \selectlanguage [0]{\@gobble}%
\providecommand \bibinfo  [0]{\@secondoftwo}%
\providecommand \bibfield  [0]{\@secondoftwo}%
\providecommand \translation [1]{[#1]}%
\providecommand \BibitemOpen [0]{}%
\providecommand \bibitemStop [0]{}%
\providecommand \bibitemNoStop [0]{.\EOS\space}%
\providecommand \EOS [0]{\spacefactor3000\relax}%
\providecommand \BibitemShut  [1]{\csname bibitem#1\endcsname}%
\let\auto@bib@innerbib\@empty
\bibitem [{\citenamefont {Bordag}\ \emph
  {et~al.}(2009{\natexlab{a}})\citenamefont {Bordag}, \citenamefont
  {Klimchitskaya}, \citenamefont {Mohideen},\ and\ \citenamefont
  {Mostepanenko}}]{book_bordag}%
  \BibitemOpen
  \bibfield  {author} {\bibinfo {author} {\bibfnamefont {M.}~\bibnamefont
  {Bordag}}, \bibinfo {author} {\bibfnamefont {G.~L.}\ \bibnamefont
  {Klimchitskaya}}, \bibinfo {author} {\bibfnamefont {U.}~\bibnamefont
  {Mohideen}}, \ and\ \bibinfo {author} {\bibfnamefont {V.~M.}\ \bibnamefont
  {Mostepanenko}},\ }\href@noop {} {\emph {\bibinfo {title} {Advances in the
  Casimir effect}}},\ Vol.\ \bibinfo {volume} {145}\ (\bibinfo  {publisher}
  {OUP Oxford},\ \bibinfo {year} {2009})\BibitemShut {NoStop}%
\bibitem [{\citenamefont {Milonni}(2013)}]{book_milonni}%
  \BibitemOpen
  \bibfield  {author} {\bibinfo {author} {\bibfnamefont {P.~W.}\ \bibnamefont
  {Milonni}},\ }\href@noop {} {\emph {\bibinfo {title} {The quantum vacuum: an
  introduction to quantum electrodynamics}}}\ (\bibinfo  {publisher} {Academic
  press},\ \bibinfo {year} {2013})\BibitemShut {NoStop}%
\bibitem [{\citenamefont {Milton}(2001)}]{book_milton}%
  \BibitemOpen
  \bibfield  {author} {\bibinfo {author} {\bibfnamefont {K.~A.}\ \bibnamefont
  {Milton}},\ }\href@noop {} {\emph {\bibinfo {title} {The Casimir effect:
  physical manifestations of zero-point energy}}}\ (\bibinfo  {publisher}
  {World Scientific},\ \bibinfo {year} {2001})\BibitemShut {NoStop}%
\bibitem [{\citenamefont {Lamoreaux}(2004)}]{lamoreaux2004casimir}%
  \BibitemOpen
  \bibfield  {author} {\bibinfo {author} {\bibfnamefont {S.~K.}\ \bibnamefont
  {Lamoreaux}},\ }\href@noop {} {\bibfield  {journal} {\bibinfo  {journal}
  {Reports on progress in Physics}\ }\textbf {\bibinfo {volume} {68}},\
  \bibinfo {pages} {201} (\bibinfo {year} {2004})}\BibitemShut {NoStop}%
\bibitem [{\citenamefont {Milton}(2004)}]{milton2004casimir}%
  \BibitemOpen
  \bibfield  {author} {\bibinfo {author} {\bibfnamefont {K.~A.}\ \bibnamefont
  {Milton}},\ }\href@noop {} {\bibfield  {journal} {\bibinfo  {journal}
  {Journal of Physics A: Mathematical and General}\ }\textbf {\bibinfo {volume}
  {37}},\ \bibinfo {pages} {R209} (\bibinfo {year} {2004})}\BibitemShut
  {NoStop}%
\bibitem [{\citenamefont {Reynaud}\ \emph {et~al.}(2001)\citenamefont
  {Reynaud}, \citenamefont {Lambrecht}, \citenamefont {Genet},\ and\
  \citenamefont {Jaekel}}]{reynaud2001quantum}%
  \BibitemOpen
  \bibfield  {author} {\bibinfo {author} {\bibfnamefont {S.}~\bibnamefont
  {Reynaud}}, \bibinfo {author} {\bibfnamefont {A.}~\bibnamefont {Lambrecht}},
  \bibinfo {author} {\bibfnamefont {C.}~\bibnamefont {Genet}}, \ and\ \bibinfo
  {author} {\bibfnamefont {M.-T.}\ \bibnamefont {Jaekel}},\ }\href@noop {}
  {\bibfield  {journal} {\bibinfo  {journal} {Comptes Rendus de l'Acad{\'e}mie
  des Sciences-Series IV-Physics}\ }\textbf {\bibinfo {volume} {2}},\ \bibinfo
  {pages} {1287} (\bibinfo {year} {2001})}\BibitemShut {NoStop}%
\bibitem [{\citenamefont {Dodonov}(2010)}]{review_dyncas}%
  \BibitemOpen
  \bibfield  {author} {\bibinfo {author} {\bibfnamefont {V.}~\bibnamefont
  {Dodonov}},\ }\href@noop {} {\bibfield  {journal} {\bibinfo  {journal}
  {Physica Scripta}\ }\textbf {\bibinfo {volume} {82}},\ \bibinfo {pages}
  {038105} (\bibinfo {year} {2010})}\BibitemShut {NoStop}%
\bibitem [{\citenamefont {Dalvit}\ \emph {et~al.}(2011)\citenamefont {Dalvit},
  \citenamefont {Neto},\ and\ \citenamefont {Mazzitelli}}]{review_friction}%
  \BibitemOpen
  \bibfield  {author} {\bibinfo {author} {\bibfnamefont {D.~A.}\ \bibnamefont
  {Dalvit}}, \bibinfo {author} {\bibfnamefont {P.~A.~M.}\ \bibnamefont {Neto}},
  \ and\ \bibinfo {author} {\bibfnamefont {F.~D.}\ \bibnamefont {Mazzitelli}},\
  }in\ \href@noop {} {\emph {\bibinfo {booktitle} {Casimir Physics}}}\
  (\bibinfo  {publisher} {Springer},\ \bibinfo {year} {2011})\ pp.\ \bibinfo
  {pages} {419--457}\BibitemShut {NoStop}%
\bibitem [{\citenamefont {Volokitin}\ and\ \citenamefont
  {Persson}(2007)}]{volokitin_persson}%
  \BibitemOpen
  \bibfield  {author} {\bibinfo {author} {\bibfnamefont {A.}~\bibnamefont
  {Volokitin}}\ and\ \bibinfo {author} {\bibfnamefont {B.~N.}\ \bibnamefont
  {Persson}},\ }\href@noop {} {\bibfield  {journal} {\bibinfo  {journal}
  {Reviews of Modern Physics}\ }\textbf {\bibinfo {volume} {79}},\ \bibinfo
  {pages} {1291} (\bibinfo {year} {2007})}\BibitemShut {NoStop}%
\bibitem [{\citenamefont {Pendry}(1997)}]{pendry97}%
  \BibitemOpen
  \bibfield  {author} {\bibinfo {author} {\bibfnamefont {J.}~\bibnamefont
  {Pendry}},\ }\href@noop {} {\bibfield  {journal} {\bibinfo  {journal}
  {Journal of Physics: Condensed Matter}\ }\textbf {\bibinfo {volume} {9}},\
  \bibinfo {pages} {10301} (\bibinfo {year} {1997})}\BibitemShut {NoStop}%
\bibitem [{\citenamefont {Pendry}(2010)}]{pendry_debate}%
  \BibitemOpen
  \bibfield  {author} {\bibinfo {author} {\bibfnamefont {J.}~\bibnamefont
  {Pendry}},\ }\href@noop {} {\bibfield  {journal} {\bibinfo  {journal} {New
  Journal of Physics}\ }\textbf {\bibinfo {volume} {12}},\ \bibinfo {pages}
  {033028} (\bibinfo {year} {2010})}\BibitemShut {NoStop}%
\bibitem [{\citenamefont {Geim}(2009)}]{rev_graphene}%
  \BibitemOpen
  \bibfield  {author} {\bibinfo {author} {\bibfnamefont {A.}~\bibnamefont
  {Geim}},\ }\href@noop {} {\bibfield  {journal} {\bibinfo  {journal} {Rev.
  Mod. Phys}\ }\textbf {\bibinfo {volume} {81}},\ \bibinfo {pages} {109}
  (\bibinfo {year} {2009})}\BibitemShut {NoStop}%
\bibitem [{\citenamefont {Kuzmenko}\ \emph {et~al.}(2008)\citenamefont
  {Kuzmenko}, \citenamefont {Van~Heumen}, \citenamefont {Carbone},\ and\
  \citenamefont {Van Der~Marel}}]{opacgraphene}%
  \BibitemOpen
  \bibfield  {author} {\bibinfo {author} {\bibfnamefont {A.}~\bibnamefont
  {Kuzmenko}}, \bibinfo {author} {\bibfnamefont {E.}~\bibnamefont
  {Van~Heumen}}, \bibinfo {author} {\bibfnamefont {F.}~\bibnamefont {Carbone}},
  \ and\ \bibinfo {author} {\bibfnamefont {D.}~\bibnamefont {Van Der~Marel}},\
  }\href@noop {} {\bibfield  {journal} {\bibinfo  {journal} {Physical review
  letters}\ }\textbf {\bibinfo {volume} {100}},\ \bibinfo {pages} {117401}
  (\bibinfo {year} {2008})}\BibitemShut {NoStop}%
\bibitem [{\citenamefont {Nair}\ \emph {et~al.}(2008)\citenamefont {Nair},
  \citenamefont {Blake}, \citenamefont {Grigorenko}, \citenamefont {Novoselov},
  \citenamefont {Booth}, \citenamefont {Stauber}, \citenamefont {Peres},\ and\
  \citenamefont {Geim}}]{opacexpgraphene}%
  \BibitemOpen
  \bibfield  {author} {\bibinfo {author} {\bibfnamefont {R.~R.}\ \bibnamefont
  {Nair}}, \bibinfo {author} {\bibfnamefont {P.}~\bibnamefont {Blake}},
  \bibinfo {author} {\bibfnamefont {A.~N.}\ \bibnamefont {Grigorenko}},
  \bibinfo {author} {\bibfnamefont {K.~S.}\ \bibnamefont {Novoselov}}, \bibinfo
  {author} {\bibfnamefont {T.~J.}\ \bibnamefont {Booth}}, \bibinfo {author}
  {\bibfnamefont {T.}~\bibnamefont {Stauber}}, \bibinfo {author} {\bibfnamefont
  {N.~M.}\ \bibnamefont {Peres}}, \ and\ \bibinfo {author} {\bibfnamefont
  {A.~K.}\ \bibnamefont {Geim}},\ }\href@noop {} {\bibfield  {journal}
  {\bibinfo  {journal} {Science}\ }\textbf {\bibinfo {volume} {320}},\ \bibinfo
  {pages} {1308} (\bibinfo {year} {2008})}\BibitemShut {NoStop}%
\bibitem [{\citenamefont {Far{\'\i}as}\ \emph {et~al.}(2013)\citenamefont
  {Far{\'\i}as}, \citenamefont {Quinteiro},\ and\ \citenamefont
  {Tamborenea}}]{farias2013}%
  \BibitemOpen
  \bibfield  {author} {\bibinfo {author} {\bibfnamefont {M.}~\bibnamefont
  {Far{\'\i}as}}, \bibinfo {author} {\bibfnamefont {G.}~\bibnamefont
  {Quinteiro}}, \ and\ \bibinfo {author} {\bibfnamefont {P.}~\bibnamefont
  {Tamborenea}},\ }\href@noop {} {\bibfield  {journal} {\bibinfo  {journal}
  {The European Physical Journal B}\ }\textbf {\bibinfo {volume} {86}},\
  \bibinfo {pages} {1} (\bibinfo {year} {2013})}\BibitemShut {NoStop}%
\bibitem [{\citenamefont {Volokitin}\ and\ \citenamefont
  {Persson}(2011)}]{volokitin2011quantum}%
  \BibitemOpen
  \bibfield  {author} {\bibinfo {author} {\bibfnamefont {A.}~\bibnamefont
  {Volokitin}}\ and\ \bibinfo {author} {\bibfnamefont {B.}~\bibnamefont
  {Persson}},\ }\href@noop {} {\bibfield  {journal} {\bibinfo  {journal}
  {Physical review letters}\ }\textbf {\bibinfo {volume} {106}},\ \bibinfo
  {pages} {094502} (\bibinfo {year} {2011})}\BibitemShut {NoStop}%
\bibitem [{\citenamefont {Far{\'\i}as}\ \emph {et~al.}(2015)\citenamefont
  {Far{\'\i}as}, \citenamefont {Fosco}, \citenamefont {Lombardo}, \citenamefont
  {Mazzitelli},\ and\ \citenamefont {L{\'o}pez}}]{farias_friction}%
  \BibitemOpen
  \bibfield  {author} {\bibinfo {author} {\bibfnamefont {M.~B.}\ \bibnamefont
  {Far{\'\i}as}}, \bibinfo {author} {\bibfnamefont {C.~D.}\ \bibnamefont
  {Fosco}}, \bibinfo {author} {\bibfnamefont {F.~C.}\ \bibnamefont {Lombardo}},
  \bibinfo {author} {\bibfnamefont {F.~D.}\ \bibnamefont {Mazzitelli}}, \ and\
  \bibinfo {author} {\bibfnamefont {A.~E.~R.}\ \bibnamefont {L{\'o}pez}},\
  }\href@noop {} {\bibfield  {journal} {\bibinfo  {journal} {Physical Review
  D}\ }\textbf {\bibinfo {volume} {91}},\ \bibinfo {pages} {105020} (\bibinfo
  {year} {2015})}\BibitemShut {NoStop}%
\bibitem [{\citenamefont {Far{\'\i}as}\ and\ \citenamefont
  {Lombardo}(2016)}]{farias2016}%
  \BibitemOpen
  \bibfield  {author} {\bibinfo {author} {\bibfnamefont {M.~B.}\ \bibnamefont
  {Far{\'\i}as}}\ and\ \bibinfo {author} {\bibfnamefont {F.~C.}\ \bibnamefont
  {Lombardo}},\ }\href@noop {} {\bibfield  {journal} {\bibinfo  {journal}
  {Physical Review D}\ }\textbf {\bibinfo {volume} {93}},\ \bibinfo {pages}
  {065035} (\bibinfo {year} {2016})}\BibitemShut {NoStop}%
\bibitem [{\citenamefont {Fosco}\ \emph {et~al.}(2007)\citenamefont {Fosco},
  \citenamefont {Lombardo},\ and\ \citenamefont
  {Mazzitelli}}]{fosco2007_moving_mirrors}%
  \BibitemOpen
  \bibfield  {author} {\bibinfo {author} {\bibfnamefont {C.}~\bibnamefont
  {Fosco}}, \bibinfo {author} {\bibfnamefont {F.}~\bibnamefont {Lombardo}}, \
  and\ \bibinfo {author} {\bibfnamefont {F.}~\bibnamefont {Mazzitelli}},\
  }\href@noop {} {\bibfield  {journal} {\bibinfo  {journal} {Physical Review
  D}\ }\textbf {\bibinfo {volume} {76}},\ \bibinfo {pages} {085007} (\bibinfo
  {year} {2007})}\BibitemShut {NoStop}%
\bibitem [{\citenamefont {Fosco}\ \emph {et~al.}(2011)\citenamefont {Fosco},
  \citenamefont {Lombardo},\ and\ \citenamefont
  {Mazzitelli}}]{fosco2011_sidewise}%
  \BibitemOpen
  \bibfield  {author} {\bibinfo {author} {\bibfnamefont {C.~D.}\ \bibnamefont
  {Fosco}}, \bibinfo {author} {\bibfnamefont {F.~C.}\ \bibnamefont {Lombardo}},
  \ and\ \bibinfo {author} {\bibfnamefont {F.~D.}\ \bibnamefont {Mazzitelli}},\
  }\href@noop {} {\bibfield  {journal} {\bibinfo  {journal} {Physical Review
  D}\ }\textbf {\bibinfo {volume} {84}},\ \bibinfo {pages} {025011} (\bibinfo
  {year} {2011})}\BibitemShut {NoStop}%
\bibitem [{\citenamefont {Fosco}\ \emph {et~al.}(2008)\citenamefont {Fosco},
  \citenamefont {Lombardo},\ and\ \citenamefont
  {Mazzitelli}}]{dynamical_matter}%
  \BibitemOpen
  \bibfield  {author} {\bibinfo {author} {\bibfnamefont {C.}~\bibnamefont
  {Fosco}}, \bibinfo {author} {\bibfnamefont {F.}~\bibnamefont {Lombardo}}, \
  and\ \bibinfo {author} {\bibfnamefont {F.}~\bibnamefont {Mazzitelli}},\
  }\href@noop {} {\bibfield  {journal} {\bibinfo  {journal} {Physics Letters
  B}\ }\textbf {\bibinfo {volume} {669}},\ \bibinfo {pages} {371} (\bibinfo
  {year} {2008})}\BibitemShut {NoStop}%
\bibitem [{\citenamefont {Maghrebi}\ \emph {et~al.}(2013)\citenamefont
  {Maghrebi}, \citenamefont {Golestanian},\ and\ \citenamefont
  {Kardar}}]{cherenkov}%
  \BibitemOpen
  \bibfield  {author} {\bibinfo {author} {\bibfnamefont {M.~F.}\ \bibnamefont
  {Maghrebi}}, \bibinfo {author} {\bibfnamefont {R.}~\bibnamefont
  {Golestanian}}, \ and\ \bibinfo {author} {\bibfnamefont {M.}~\bibnamefont
  {Kardar}},\ }\href@noop {} {\bibfield  {journal} {\bibinfo  {journal}
  {Physical Review A}\ }\textbf {\bibinfo {volume} {88}},\ \bibinfo {pages}
  {042509} (\bibinfo {year} {2013})}\BibitemShut {NoStop}%
\bibitem [{\citenamefont {Bordag}\ \emph
  {et~al.}(2009{\natexlab{b}})\citenamefont {Bordag}, \citenamefont
  {Fialkovsky}, \citenamefont {Gitman},\ and\ \citenamefont
  {Vassilevich}}]{bordag_casimir_graphene}%
  \BibitemOpen
  \bibfield  {author} {\bibinfo {author} {\bibfnamefont {M.}~\bibnamefont
  {Bordag}}, \bibinfo {author} {\bibfnamefont {I.}~\bibnamefont {Fialkovsky}},
  \bibinfo {author} {\bibfnamefont {D.}~\bibnamefont {Gitman}}, \ and\ \bibinfo
  {author} {\bibfnamefont {D.}~\bibnamefont {Vassilevich}},\ }\href@noop {}
  {\bibfield  {journal} {\bibinfo  {journal} {Physical Review B}\ }\textbf
  {\bibinfo {volume} {80}},\ \bibinfo {pages} {245406} (\bibinfo {year}
  {2009}{\natexlab{b}})}\BibitemShut {NoStop}%
\bibitem [{\citenamefont {Fialkovsky}\ \emph {et~al.}(2011)\citenamefont
  {Fialkovsky}, \citenamefont {Marachevsky},\ and\ \citenamefont
  {Vassilevich}}]{fialkovsky2011finite}%
  \BibitemOpen
  \bibfield  {author} {\bibinfo {author} {\bibfnamefont {I.~V.}\ \bibnamefont
  {Fialkovsky}}, \bibinfo {author} {\bibfnamefont {V.~N.}\ \bibnamefont
  {Marachevsky}}, \ and\ \bibinfo {author} {\bibfnamefont {D.~V.}\ \bibnamefont
  {Vassilevich}},\ }\href@noop {} {\bibfield  {journal} {\bibinfo  {journal}
  {Physical Review B}\ }\textbf {\bibinfo {volume} {84}},\ \bibinfo {pages}
  {035446} (\bibinfo {year} {2011})}\BibitemShut {NoStop}%
\bibitem [{\citenamefont {Bordag}\ \emph {et~al.}(2016)\citenamefont {Bordag},
  \citenamefont {Fialkovskiy},\ and\ \citenamefont
  {Vassilevich}}]{bordag2016enhanced}%
  \BibitemOpen
  \bibfield  {author} {\bibinfo {author} {\bibfnamefont {M.}~\bibnamefont
  {Bordag}}, \bibinfo {author} {\bibfnamefont {I.}~\bibnamefont {Fialkovskiy}},
  \ and\ \bibinfo {author} {\bibfnamefont {D.}~\bibnamefont {Vassilevich}},\
  }\href@noop {} {\bibfield  {journal} {\bibinfo  {journal} {Physical Review
  B}\ }\textbf {\bibinfo {volume} {93}},\ \bibinfo {pages} {075414} (\bibinfo
  {year} {2016})}\BibitemShut {NoStop}%
\end{thebibliography}

%

\end{document}